# Single-pulse-based Interstellar Scintillation Studies of RRATs

Zi-wei Wu[1], Wei-wei Zhu[1,2], Zi-yao Fang[1], Qiu-yang Fu[1,3], Ji-guang Lu[1,4,5], Ling-qi Meng[1,3], Chen-Chen Miao[6], Xue-li Miao[1], Jia-rui Niu[1], Rukeya Rejiefu[1], Xun Shi[7], Chao Wang[1], Meng-yao Xue[1], Mao Yuan[8], You-ling Yue[1], Chun-feng Zhang[1], Zhen Zhang[1], Shi-jun Dang[5,9,10], and Yu-lan Liu[1,4,5]

[1] National Astronomical Observatories, Chinese Academy of Sciences, Beijing 100101, People's Republic of China; wuzw@bao.ac.cn, zhuww@nao.cas.cn
[2] Institute for Frontier in Astronomy and Astrophysics, Beijing Normal University, Beijing 102206, People's Republic of China
[3] School of Astronomy and Space Science, University of Chinese Academy of Sciences, Beijing 100049, People's Republic of China
[4] CAS Key Laboratory of FAST, National Astronomical Observatories, Chinese Academy of Sciences, Beijing 100101, People's Republic of China
[5] Guizhou Radio Astronomical Observatory, Guizhou University, Guiyang 550001, People's Republic of China
[6] Zhejiang Lab, Kechuang Avenue, Zhongtai Sub-District, Yuhang District, Hangzhou, Zhejiang Province, People's Republic of China
[7] South-Western Institute for Astronomy Research (SWIFAR), Yunnan University, Kunming 650500, People's Republic of China
[8] National Space Science Center, Chinese Academy of Sciences, Beijing 100190, People's Republic of China
[9] School of Physics and Electronic Science, Guizhou Normal University, Guiyang 550025, People's Republic of China
[10] Guizhou Provincial Key Laboratory of Radio Astronomy and Data Processing, Guiyang 550025, People's Republic of China
Received 2025 January 22; revised 2025 March 17; accepted 2025 March 19; published 2025 March 28

## Abstract

The nature of irregularly spaced pulses of rotating radio transients (RRATs) complicates interstellar scintillation studies. In this Letter, we report the primary scintillation parameters of a sample of RRATs using pairwise correlations of pulse spectra. Moreover, from the measured scintillation velocities, we constrain their transverse velocities. We also find a reduced modulation index, $m = 0.13 \pm 0.01$, for RRAT J1538+2345. Several possible explanations are discussed. Furthermore, the single-pulse-based interstellar scintillation technique is applicable to other pulsar populations, including nulling pulsars and those with short scintillation timescales, and fast radio bursts.

*Unified Astronomy Thesaurus concepts:* Radio transient sources (2008); Interstellar scintillation (855)

## 1. Introduction

Pulsars (A. Hewish et al. 1968), rotating radio transients (RRATs; M. A. McLaughlin et al. 2006), and fast radio bursts (FRBs; D. R. Lorimer et al. 2007) have been discovered through their folded pulse profiles or single pulses. Pulsars are magnetized neutron stars (F. Pacini 1967, 1968; M. I. Large et al. 1968; P. B. Demorest et al. 2010). The genesis of FRBs remains a contentious issue (B. Zhang 2020). However, the link between a FRB and a highly magnetized pulsar has been posited (C. D. Bochenek et al. 2020; CHIME/FRB Collaboration et al. 2020; L. Lin et al. 2020; C. K. Li et al. 2021; W. Zhu et al. 2023). More recent phenomena detected from FRBs (J. R. Niu et al. 2024; R. Mckinven et al. 2025; K. Nimmo et al. 2025) also support that at least some FRBs originate from pulsars. RRATs are a subclass of pulsars characterized by sporadic single-pulse radiation. RRATs could be very nulling pulsars (D. J. Zhou et al. 2023), indicating they are characteristic of old age (R. T. Ritchings 1976). However, a glitch from RRAT J1819−1458 suggests a young pulsar or magnetar origin (A. G. Lyne et al. 2009).

The beams of radiation emitted from these compact radio sources are subject to being perturbed due to fluctuations in the refractive index of the ionized interstellar medium (IISM). These perturbations induce phase changes across the disparate rays of light. The interference of these inherently uncorrelated scattered rays leads to a modulation of the pulse intensity as a function of both frequency and time, a phenomenon widely recognized as interstellar scintillation (ISS; P. A. G. Scheuer 1968). The past nearly five decades have shown increasing interest and applications of pulsar ISS, i.e., constraining proper motions (A. G. Lyne & F. G. Smith 1982), mass measurements (A. G. Lyne 1984), resolving pulsar emission (R. Main et al. 2018), and 3D spin-velocity alignment (J. Yao et al. 2021).

The previous pulsar ISS works are typically performed by averaging several adjacent pulses. RRATs are also scintillating sources. However, in contrast to typical pulsars, FRBs, and RRATs emit radio radiations with irregularly spaced bursts or pulses, complicating the study of their scintillation characteristics (B. W. Meyers et al. 2019; J. Xie et al. 2022). Moreover, the intrinsic variability of pulsar emissions, coupled with the limited signal-to-noise ratio of the single pulses, complicates single-pulse-based studies of ISS. However, recent works have enabled detailed investigation of the scintillation properties of FRBs (R. A. Main et al. 2022; Z.-W. Wu et al. 2024; Z. Wu et al. 2024), and Crab giant pulses (J. M. Cordes et al. 2004; R. Main et al. 2021; R. Lin et al. 2023) with the consideration of minimizing the noise from intrinsic variations in radio emission. Consequently, it is plausible to extend the application of the methodologies developed for FRBs to conduct scintillation studies on RRATs.

In this Letter, we elucidate the scintillation characteristics of six RRATs, as observed through the Five-hundred-meter Aperture Spherical radio Telescope (FAST). The observations and data processing procedures are delineated in Section 2. Section 3 is dedicated to the results and discussions. The summation of our conclusions is encapsulated in Section 4.

## 2. Observations and Data Analysis

The received radio emissions of known RRATs are weak. The mean value of their flux densities is 0.13 mJy at 1400 MHz

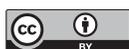

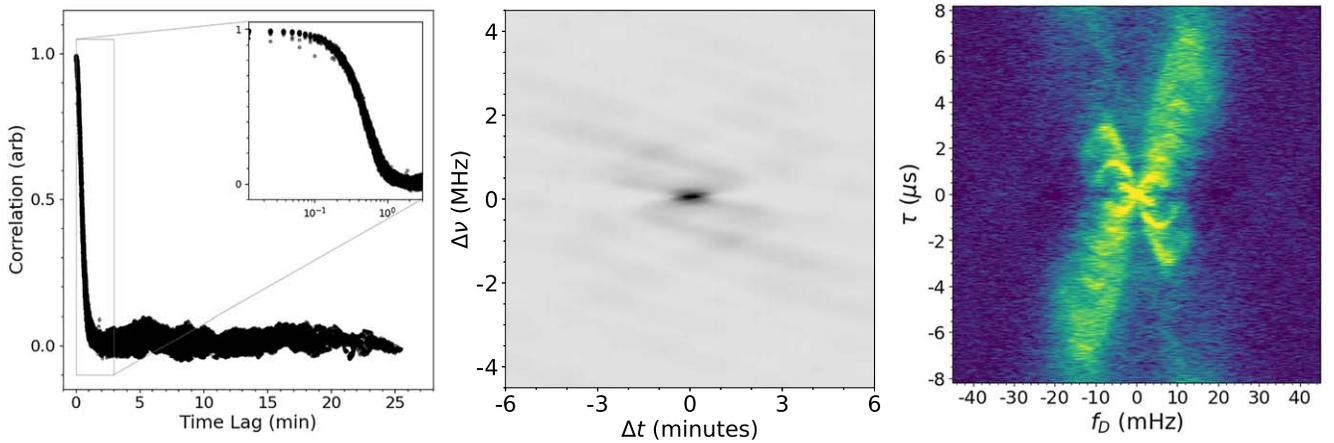

**Figure 1.** Measured correlation coefficients between single-pulse pairs of pulsar J1509+5531 (left panel), the corresponding 2D ACF (middle panel), and secondary spectrum (right panel) with FAST at 1.0–1.5 GHz with frequency resolution of 61 kHz. We select 271 bright single pulses (the peak signal-to-noise ratio ⩾ 500) spanning about 26 minutes to do the experiment of single-pulse-based interstellar scintillation. The number of correlation coefficient measurements is larger than 73,000. The time resolution of the time lag of correlation coefficients is the spin period of this pulsar (∼0.74 s). The inverted parabolic arclets are visible. Note that the axis scales of the 2D ACF and secondary spectrum are adjusted to highlight the majority of the details.

**Table 1**
Summary of FAST Observations and Scintillation Properties of Selected RRATs

| RRATs | DM (cm$^{-3}$ pc) | $P$ (s) | Time (MJD) | $T_{\rm obs}$ (minutes) | $T_{\rm obs}/P$ | $N_{\rm p}$ | $\Delta\nu_{\rm d}$ (MHz) | $\tau_{\rm d}$ (minutes) | $V_{\rm T}$ (km s$^{-1}$) | Note |
|---|---|---|---|---|---|---|---|---|---|---|
| J0103+54 | 55.6 | 0.4 | 59512 | 60 | 10,160 | 0 | ... | ... | ... | No pulse |
| J0139+3336 | 21.2 | 1.2 | 58865 | 200 | 9615 | 110 | 4.7 ± 0.4 | 6.5 ± 0.2 | 138 ± 20 | ... |
| J0628+0909 | 88.3 | 1.2 | 59511 | 64 | 3093 | 162 | <0.122 | 0.72 ± 0.02 | <226 ± 30 | ... |
| J1538+2345 | 14.9 | 3.4 | 58338 | 30 | 521 | 159 | 13.4 ± 2.2 | 11.1 ± 1.8 | 120 ± 30 | ... |
| J1854+0306 | 192.4 | 4.6 | 58842 | 63 | 829 | 162 | ... | ... | ... | limited Δf |
| J1913+1330 | 175.6 | 0.9 | 58833 | 182 | 11,858 | 555 | <0.122 | 0.34 ± 0.02 | <870 ± 100 | ... |

**Note.** Given are the RRATs name, dispersion measure, and spin period; along with the observation date, the observation duration, the length of the observation in the unit of the spin period, the bright burst number we use for scintillation studies, the derived scintillation parameters over the range of 1.0–1.5 GHz assuming a halfway location, stationary and isotropic screen, and a note on the likely cause for our nondetections of scintillation.

from psrcat[11] (V2.5.1; R. N. Manchester et al. 2005). Consequently, the sensitivity of the radio telescope is pivotal for the detection of RRATs and enhances its signal-to-noise. FAST operates with an effective illuminated aperture of approximately 300 meters (R. Nan et al. 2011), rendering it an exemplary instrument for conducting studies on the scintillation of RRATs. Our observations are conducted using the central beam of a 19 beam receiver, featuring a bandwidth that spans from 1.0 to 1.5 GHz, and a digital backend based on Reconfigurable Open-architecture Computing Hardware-version2 (D. Li et al. 2018; P. Jiang et al. 2019, 2020). The data are logged in the search mode using the PSRFITS format (A. W. Hotan et al. 2004). The pulsar ephemeris is sourced from psrcat. We record data across four polarizations, with 4096 phase bins and 4096 frequency channels, corresponding to a frequency resolution ($\Delta f$) of 0.122 MHz. Frequency channels with significant radio interference are subsequently identified and manually excised.

We curate a selection of six RRATs, chosen based on their pulse rate as documented in the RRATalog[12] and the duration of FAST observations, to ensure a sufficient accumulation of pulses for analysis. A summary of the selected RRATs, including the name, dispersion measure (DM), spin period ($P$), MJD, the actual observation duration ($T_{\rm obs}$), and the ratio of observation length to spin period ($T_{\rm obs}/P$), is presented in Table 1. We exclusively consider pulses with a peak signal-to-noise ratio (S/N)[13] exceeding a threshold of 10.0. This criterion is essential as scintillation studies often rely on signals from rays that are slightly deflected, typically accounting for less than 1% of the pulse's mean flux. The methodologies for constructing dynamic and secondary spectra are outlined in prior works by R. A. Main et al. (2022) and Z.-W. Wu et al. (2024). We use mean-normalized spectral autocovariance for modulation index analysis (J. P. Macquart et al. 2019). The uncertainty associated with the scintillation bandwidth $\Delta\nu_{\rm d}$ and scintillation timescale $\tau_{\rm d}$ is determined by the quadrature sum of the uncertainties derived from the fitting process and the statistical error $\sigma_{\rm est}$, which arises from the finite number of scintles observed (i.e., N. D. R. Bhat et al. 1999):

$$\sigma_{\rm est} = \left(f_d \times \frac{{\rm BW}_{\rm obs} T_{\rm obs}}{\Delta\nu_{\rm d} \tau_{\rm d}}\right)^{-0.5}, \quad (1)$$

where BW$_{\rm obs}$ is the observing frequency bandwidth, and $f_d$ (=0.4) is the filling factor.

We show the correlation coefficients between pulse pairs of PSR J1509+5531 ($\Delta f \sim 0.061$ MHz) at FAST and the corresponding secondary spectrum in Figure 1. With the single-pulse-based ISS technique, we measure the scintillation bandwidth and timescale to be $100 \pm 6$ kHz and $29 \pm 1$ s from 2D autocorrelation function (ACF; Figure 1), respectively. In comparison, they are inferred to be $128 \pm 9$ kHz and $33 \pm 1$ s from averaged-adjacent-pulses ISS (five single pulses; Z. Wu et al. 2022). We utilize only approximately 15% of single pulses for conducting our single-pulse-based ISS experiment, which should account for the minor discrepancies observed in scintillation parameters between the two methodologies. Our results strongly validate the potential of single-pulse-based ISS. A similar analysis on a large sample of normal pulsars will be presented in another paper.

---

[11] https://www.atnf.csiro.au/research/pulsar/psrcat/
[12] https://rratalog.github.io/rratalog/
[13] The peak S/N is given by the ratio of the maximum intensity of the pulse profile to the standard deviation of the off-pulse region.

### 3. Results and Discussions

In our FAST observation, no pulses from RRAT J0103+54 at 1250 MHz are detected. This RRAT was initially identified by the Green Bank Telescope at 350 MHz, exhibiting a pulse rate of $390 \pm 108$ hr$^{-1}$ (C. Karako-Argaman et al. 2015). However, the aforementioned study did not find pulses at the elevated frequency of 820 MHz. This could be due to frequency-dependent or temporal variations in the activity of RRAT J0103+54.

The number of bright pulses ($N_p$, as detailed in Table 1) selected for our analysis is fewer than in previous studies on the same data set, i.e., J0139+3336 (S. J. Dang et al. 2024), J1538+2345 (J. Lu et al. 2019), J1854+0306 (Q. Guo et al. 2024), and J1913+1330 (S. B. Zhang et al. 2024), due to the application of a more stringent detection threshold.

Subsequently, we derive the primary scintillation parameters, scintillation timescale $\tau_d$, and scintillation bandwidth $\Delta\nu_d$ for the remaining RRATs, as presented in Table 1, utilizing the 2D ACF. For RRAT J1854+0306 we cannot resolve its scintillation bandwidth $\Delta\nu_d$ at all. It is anticipated to be approximately 1 kHz from the NE2001 galactic electron density model (J. M. Cordes & T. J. W. Lazio 2002), a value significantly lower than our frequency resolution $\Delta f$ of 0.122 MHz. We depict the 2D ACFs in Figure 2.

#### 3.1. Constraining the Transverse Velocity

Pulsars, or neutron stars, are characterized by their high transverse velocities (B. Hansen & E. S. Phinney 1997; G. Hobbs et al. 2005). These velocities are hypothesized to arise from asymmetric supernova explosions, which impart a kick velocity to the nascent neutron star (R. J. Dewey & J. M. Cordes 1987).

The transverse velocity of pulsars is typically inferred from extensive pulsar timing studies, multiepoch interferometric observations, optical telescopic data, and ISS. To date, these methodologies have not been successfully adapted for the measurement of the transverse velocity of RRATs. This limitation arises due to the distinct observational characteristics of RRATs; for example, whereas pulsars are generally timed using their average pulse profiles, RRATs require timing from their sporadic individual pulses. Even when timing solutions are ascertained, their uncertainties tend to be substantially greater than those for regular pulsars. This disparity is attributed to pulse-to-pulse jitter noise (A. G. Lyne et al. 2009) and complicated pulse-profile morphology (S. J. Dang et al. 2024; S. B. Zhang et al. 2024).

The transverse velocity of RRATs is correlated with the scintillation velocity (A. G. Lyne & F. G. Smith 1982; J. M. Cordes 1986) in the form of (J. M. Cordes & B. J. Rickett 1998):

$$V_{\text{eff}}(s) = (1-s)(V_T + V_O) + sV_E - V_{\text{IISM}}(s), \qquad (2)$$

where $V_T$, $V_O$, $V_E$, and $V_{\text{IISM}}$ are the velocities of the RRAT proper motion, the possible orbital motion of the RRAT, earth motion, and the IISM, respectively. In essence, it should be feasible to deduce constraints on the transverse velocity of RRATs by examining ISS effects, with data from a single epoch. The distance of these RRATs is provided by YMW16 (J. M. Yao et al. 2017) with 25% uncertainty included. Under the assumption of a halfway location, stationary and isotropic screen, and subtracting the earth velocity, we constrain the transverse velocities of RRATs. The resultant transverse velocities $V_T$ span a range from 120 to 870 km s$^{-1}$, as detailed in Table 1. These derived velocities of RRATs are notably greater compared to those of ordinary stars (Gaia Collaboration et al. 2018). They are indeed comparable to those of pulsars (G. Hobbs et al. 2005).

#### 3.2. Interpreting Modulation Indices

The quantitative nature of the transverse separation of pulsar emission region has remained enigmatic. ISS affords the potential to resolve the tiny emission regions of pulsars since IISM can act as a lens, which has the ability to achieve picoarsecond angular resolution (M. D. Johnson et al. 2012; U. L. Pen et al. 2014).

The basic idea is that the size of the emission region affects the distribution of ISS-caused flux variation quantified by modulation index ($m$; E. E. Salpeter 1967), and an angular displacement in the emission site causes a lateral displacement in the scintillation pattern at Earth. Various attempts have tried to measure gates or pulse components dependence of ISS (i.e., J. M. Cordes et al. 1983; A. Wolszczan & J. M. Cordes 1987; C. R. Gwinn et al. 2012). There are attempts to measure the modulation index by the distribution function of flux intensity (C. R. Gwinn et al. 1997) and ACF (J. P. Macquart et al. 2000). For a point source $m = 1$; for an extended source $m < 1$, with a smaller modulation index $m$ for a larger radiation region diameter of source, which is exactly the reason "stars twinkle, planets do not." Modulation index has been a routine method to infer source sizes in all types of scintillation effects, for example, interplanetary scintillation resulting from solar wind (A. Salmona 1967; A. Hewish et al. 1974; W. A. Coles & J. J. Kaufman 1977).

To sufficiently sample ISS-caused flux variation from a finite number and small size of scintles, an observation well suited for modulation index studies should have an observing bandwidth $B_{\text{obs}} \gg \Delta\nu_d \gg \Delta f$, and an observation duration $T_{\text{obs}} \gg \tau_d \gg P \gg \Delta t$ (sampling time). Thus, single-pulse-based ISS can better resolve the flux variation if we have a high-quality (S/N) pulse profile and also minimize the intrinsic variation in pulsar emission. In our cases of four RRATs, the limited frequency resolution of J0628+0909 and J1913+1330 can reduce the modulation index. The distribution of ACF of J0139+3336 is scattered because of the heavy radio frequency interference contamination. RRAT J1538+2345 is the most suitable source for the purpose of modulation index studies. It shows a smaller modulation index $m \approx 0.13 \pm 0.01$ compared to that of $m \approx 0.99 \pm 0.01$ of J1509+5531 (Figure 1). It is also significantly smaller than the one-third correlation expected from a randomized signal imparted with the same impulse response function (J. M. Cordes et al. 2004), underconsidering the polarization of the substructure (R. Lin et al. 2023).

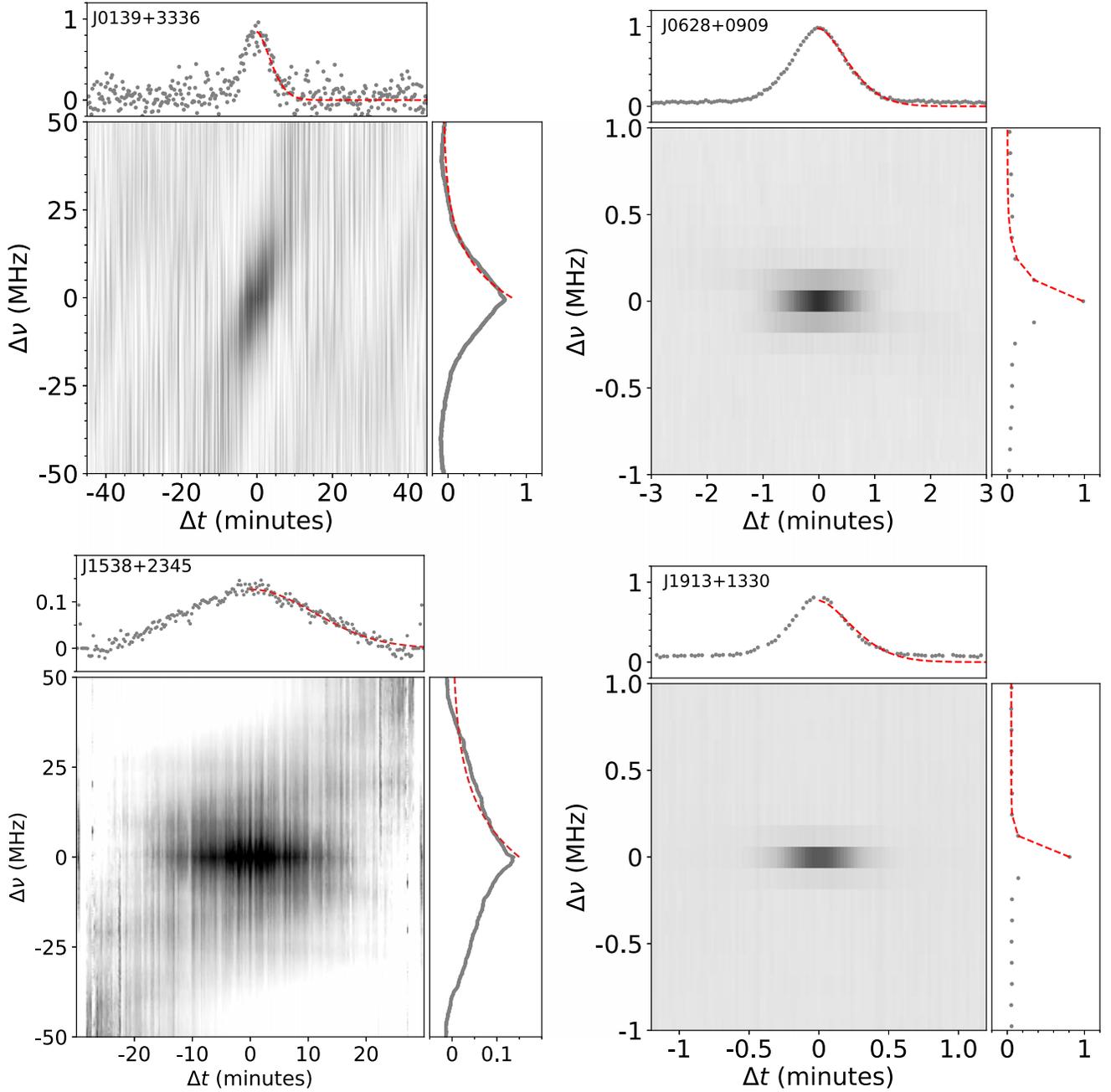

**Figure 2.** The 2D ACFs of four RRATs with FAST. In the two smaller side plots, the gray points are the 1D ACFs at zero frequency and time lag, and the red dashed curves are the best fits from which the scintillation timescale $\tau_d$ and scintillation bandwidth $\Delta\nu_d$ are derived, respectively.

Assuming a halfway location of the scattering screen of J1538+2345, the physical spatial resolution of the screen $\Delta x$ is about $4.6 \pm 0.7 \times 10^4$ km. The derived transverse separation of emission region $R_e$ is about $1.7 \pm 0.3 \times 10^5$ km, which is comparable to light-cylinder radius $R_{LC} \equiv cP/2\pi = 1.6 \times 10^5$ km. The resolution of the scattering screen is smaller than its light-cylinder radius. A closer scattering screen near the pulsar itself would always provide a much higher spatial resolution, i.e., $d_p - d_s = 2$ pc, $\Delta x \approx 1800$ km. It looks like a reduced modulation index of RRAT J1538+2345 is caused by a spatially extended source. In this instance, a resolved source should have a modulation index increasing with increasing frequency (C. R. Gwinn et al. 1998). However, further investigation finds that the modulation index seems to present opposite scaling, which can be seen from pulsars in weak scintillation regimes since phase fluctuations get weaker at higher frequencies (B. J. Rickett 1990). The evolution trend of the modulation index is hard to identify due to the limited frequency coverage of our data set. Future ultrawide bandwidth observations are needed.

There are also other possibilities for generating a reduced modulation index. In the presence of a multiple-screen scattering system (C. R. Gwinn et al. 2006), a screen with unresolved scintillation bandwidth in the current frequency resolution would wash out the observed scintillation

(C. R. Gwinn et al. 1998). The short noise with the pulse and birefringence causing an averaging of scintillation patterns between left and right circular polarization would also moderately affect the modulation index ($m \geqslant 1/3$; R. Lin et al. 2023).

## 4. Conclusions

In this Letter, we apply the single-pulse-based interstellar scintillation on RRATs at 1.25 GHz with FAST, successfully testing their scintillation properties. Furthermore, we constrain transverse velocities $V_T$ of these RRATs from the perspective of scintillation velocity, assuming a halfway location and stationary scattering screen. The derived values of $V_T$ ranging from 120 to 870 km s$^{-1}$ are relatively large compared to normal stars (Gaia Collaboration et al. 2018). Yet they are consistent with another catalog of pulsars. We detect a reduced modulation index $m = 0.13 \pm 0.01$ from RRAT J1538+2345.

We also conduct a search for scintillation arcs (J. M. Cordes et al. 2006; U.-L. Pen & Y. Levin 2014) in RRATs, a phenomenon observed in pulsars (D. R. Stinebring et al. 2001; Z. Wu et al. 2022; R. A. Main et al. 2023) and two FRBs (Z. Wu et al. 2024; Z.-W. Wu et al. 2024). No distinct parabolic structures are discerned within the secondary spectra of RRATs, which could simply be a resolution issue. It is also possible that we could see scintillation arcs from these selected RRATs in the future due to a change of images (T. Sprenger et al. 2022).

Multiple-pulses averaged ISS has been a routine method of deriving the transverse velocity of pulsars and the emission size of radio sources. Here, we demonstrate that the single-pulse-based ISS also presents a promising ISS method for normal pulsars and pulsars with irregularly spaced pulses. Fully quantifying the transverse separation of emission size is yet impossible, the uncertainties come from the location and anisotropic of the scattering screen and the distance of the pulsar. Our forthcoming research endeavors will encompass the measurement of additional proper motions through the ISS technique and confirm some of them with interferometry or pulsar timing. It is anticipated that the determination of vector transverse velocities may be achievable by conducting ISS measurements across an entire annual cycle, similar to typical pulsars (Y. Liu et al. 2023).

With the advancements of telescopes in recent years, there has been a proliferation of surveys dedicated to the discovery of RRATs (F. A. Dong et al. 2023; W. Fiore et al. 2023; D. J. Zhou et al. 2023), resulting in a swift increase in the number of RRATs detected. We advocate for expanding the RRAT scintillation census to acquire a more extensive data set of scintillation velocities and search for scintillation arcs. This could shed new light on the pulsar ISS applications, such as conducting a systematic comparison in transverse velocities between different types of pulsars, resolving pulsar emission regions, probing the properties of the IISM, and investigating the geometry and kinematic behaviors of the pulsars. Furthermore, long-term monitoring of highly active RRATs is essential for the investigation of their variability (e.g., A. G. Lyne et al. 2009; T. V. Smirnova et al. 2022).

## Acknowledgments

We thank the anonymous referee for the constructive comments and suggestions, which helped us to improve this paper. This work is supported by the National SKA Program of China Nos. 2020SKA0120200 and 2020SKA0120100, CAS Project for Young Scientists in Basic Research YSBR-063, the National Nature Science Foundation grant Nos. 12041303, 11988101, 11833009, 11873067, 12041304, 12273009, 12203045, 12203013, and 12303042, and the National Key R&D Program of China Nos. 2017YFA0402600, 2021YFA0718500, 2017YFA0402602, and 2022YFC2205203, the CAS-MPG LEGACY project, the Max-Planck Partner Group, the Chinese Academy of Sciences Project for National Science Foundation, China, No. 12421003, the Strategic Priority Research Program of the Chinese Academy of Sciences (No. XDA0350501), and the Major Science and Technology Program of Xinjiang Uygur Autonomous Region, grant No. 2022A03013-2. Z.W. is supported by the Project funded by China Postdoctoral Science Foundation No. 2023M743517. S.D. is supported by Guizhou Provincial Science and Technology Foundation (No. ZK[2022]304), the Major Science and Technology Program of Xinjiang Uygur Autonomous Region (No. 2022A03013-4), the Scientific Research Project of the Guizhou Provincial Education (No. KY[2022]132) and Foundation of Education Bureau of Guizhou Province, China (grant No. KY (2020) 003). This work made use of the data from Five-hundred-meter Aperture Spherical radio Telescope (FAST) (https://cstr.cn/31116.02.FAST). FAST is a Chinese national mega-science facility, operated by National Astronomical Observatories, Chinese Academy of Sciences.

## Data Availability

The FAST data center already published raw data that can be accessed there.

## ORCID iDs

Zi-wei Wu ● https://orcid.org/0000-0002-1381-7859
Wei-wei Zhu ● https://orcid.org/0000-0001-5105-4058
Ji-guang Lu ● https://orcid.org/0000-0002-9815-5873
Ling-qi Meng ● https://orcid.org/0000-0002-2885-568X
Chen-Chen Miao ● https://orcid.org/0000-0002-9441-2190
Xue-li Miao ● https://orcid.org/0000-0003-1185-8937
Jia-rui Niu ● https://orcid.org/0000-0001-8065-4191
Rukeya Rejiefu ● https://orcid.org/0000-0002-3283-075X
Xun Shi ● https://orcid.org/0000-0003-2076-4510
Meng-yao Xue ● https://orcid.org/0000-0001-8018-1830
Mao Yuan ● https://orcid.org/0000-0003-1874-0800
You-ling Yue ● https://orcid.org/0000-0003-4415-2148
Chun-feng Zhang ● https://orcid.org/0000-0002-4327-711X
Shi-jun Dang ● https://orcid.org/0000-0002-2060-5539
Yu-lan Liu ● https://orcid.org/0000-0001-9986-9360